\documentclass[draftcls,onecolumn,peerreview]{IEEEtran}
\newcommand{\qed}{\nobreak \ifvmode \relax \else
      \ifdim\lastskip<1.5em \hskip-\lastskip
      \hskip1.5em plus0em minus0.5em \fi \nobreak
      \vrule height0.75em width0.5em depth0.25em\fi}

\ifCLASSINFOpdf
  \usepackage[pdftex]{graphicx}
  \graphicspath{{./}{Figs/}}
  \DeclareGraphicsExtensions{.pdf,.jpeg,.png}
\else
\fi
\usepackage{amssymb}
\usepackage{amsmath}
\usepackage{mathtools}
\usepackage{bigdelim}

\begin{document}
\title{Defeating the Eavesdropper: On the Achievable Secrecy Capacity using Reconfigurable Antennas}

\author{Ahmed M. Alaa

\thanks{The authors are with XX}
\thanks{Manuscript received XXXX XX, 2013; revised XXXX XX, 201X.}}

\markboth{XXXXXX}%
{A. M. Alaa \MakeLowercase{\textit{et al.}}: Secret Selection Diversity using Confidential Reconfigurable Antenna State Switching Patterns}

\maketitle
\begin{abstract}
In this paper, we consider the transmission of confidential messages over slow fading wireless channels in the presence of an eavesdropper. We propose a transmission scheme that employs a single reconfigurable antenna at each of the legitimate partners, whereas the eavesdropper uses a single conventional antenna. A reconfigurable antenna can switch its propagation characteristics over time and thus it perceives different fading channels. It is shown that without channel side information (CSI) at the legitimate partners, the main channel can be transformed into an ergodic regime offering a \textit{secrecy capacity} gain for strict outage constraints. If the legitimate partners have partial or full channel side information (CSI), a sort of selection diversity can be applied boosting the maximum secret communication rate. In this case, fading acts as a friend not a foe.  
\end{abstract}


\IEEEpeerreviewmaketitle

\section{Introduction}
\IEEEPARstart{I}{nformation} theoretic security was quantified by Shannon's notion of \textit{perfect secrecy}. Perfect information-theoretic secrecy requires the signal received by the eavesdropper not to provide any additional information about the transmitted message \cite{IEEEhowto:kopka1}. The conventional secret communications scheme includes two legitimate parties, commonly known as Alice and Bob, communicating over a wireless slow fading channel. A malicious third party, known as Eve, eavesdrops on the wireless medium and tries to decode the transmitted signal. In a block fading channel, the channel gain is constant over a codeword, thus the channel is characterized by an outage behavior. The achievable secrecy rate was obtained in terms of the outage probability in \cite{IEEEhowto:kopka2}, where it is shown that for a fading channel, poor secret rates are achieved for strict outage constraints.

Recently, improving the outage secrecy capacity by using multiple antennas has been studied \cite{IEEEhowto:kopka11} \cite{IEEEhowto:kopka12} \cite{IEEEhowto:kopka13}. However, the usage of multiple antennas is inhibited by the space limitations in many wireless transceivers. In addition to that, multiple antennas require multiple RF chains which increases the cost and complexity of the wireless transceiver. In this work, we propose a novel secret communications scheme that employs \textit{reconfigurable antennas}; a class of antennas capable of changing one of its characteristics (polarization, operating frequency and radiation pattern) over time \cite{IEEEhowto:kopka3} \cite{IEEEhowto:kopka5} using a single RF chain. Each configuration is known as a \textit{radiation state} and corresponds to an independent channel realization. Previous research work utilized reconfigurable antennas in authentication and secret key generation \cite{IEEEhowto:kopka15} \cite{IEEEhowto:kopka17}. However, the achievable capacity bounds for reconfigurable antenna schemes were never obtained before. We propose two modes of legitimate communication via reconfigurable antennas: \textit{state switching} and \textit{state selection}. State switching is applied by the CSI is not available at the transmitter/receiver and relies on switching the antenna \textit{radiation state} over time manipulating the wireless channel and creating artificial channel fluctuations. On the other hand, state selection is applied by selecting the ``best" radiation state per codeword for a block fading channel based on the CSI at the transmitter/receiver. It is shown that when strict outage constraints are imposed on the system, state switching can offer an ergodic capacity that exceeds the achievable outage capacity. Moreover, state selection based on partial or full CSI can offer a secrecy capacity that exceeds that of the AWGN channel, thus fading acts as a friend not a foe. State selection resembles opportunistic transmission in a fast fading channel but with power allocation in the \textit{state domain} rather than the time domain, thus supporting both \textit{delay constrained} and \textit{delay tolerant} applications.   

As shown in figure 1, we modify the conventional secrecy communications scheme by employing a reconfigurable antenna at both of the legitimate parties. A message $W^k$ is mapped to a codeword $X^n$. The codeword is then transmitted from Alice to Bob via a rayleigh fading channel $\gamma^n_{M} = \{\gamma^n_{M}(1), \ldots, \gamma^n_{M}(n)\}$, and Additive White Gaussian Noise (AWGN) = $\{n^n_{M}(1), \ldots, n^n_{M}(n)\}$, where $n^n_{M}(i)$ $\sim$ $\mathcal{CN}(0, 1)$. The estimated message by the decoder is obtained by demapping the received signal $Y^n_{M}$ to $\hat{W}^k$. Eve, an eavesdropper, receives the signal via a similar channel $\gamma^n_{W}$, and noise $n^n_{W}$. While Eve uses a conventional single antenna, both Bob and Alice use reconfigurable antennas with $Q_{R}$ and $Q_{T}$ propagation modes respectively. The realizations $\gamma^n_{M}(i)$ and $\gamma^n_{W}(i)$ are the legitimate and eavesdropper channel realizations for the $i^{th}$ symbol within a codeword of length $n$. For a slow fading channel, both are constant over a codeword. However, a reconfigurable antenna is capable of switching the channel state once per symbol, thus there are $Q_{R}Q_{T}$ possible realizations for the main channel $\big(\gamma^n_{M}(i) \in \{\gamma_{M}(1),\ldots,\gamma_{M}(Q_{R}Q_{T})\}\big)$ and $Q_{T}$ possible realizations for the eavesdropper channel. Note that all channels are assumed to be Rayleigh fading channels, thus the \textit{probability density functions} (pdfs) of the channels are $f_{\gamma}(\gamma_{M}) = \frac{1}{\overline{\gamma}_{M}} e^{-\frac{\gamma_{M}}{\overline{\gamma}_{M}}}$ and $f_{\gamma}(\gamma_{W}) = \frac{1}{\overline{\gamma}_{W}} e^{-\frac{\gamma_{W}}{\overline{\gamma}_{W}}}$, where $\overline{\gamma}_{W}$ and $\overline{\gamma}_{M}$ are the average SNR values for the main and eavesdropper channels. We define the error probability $P_{e}^n$ as the average probability of erroneous decoding of the received message, and the equivocation rate $R_{e}$ as the entropy rate of the transmitted message conditioned on the channel outputs at the eavesdropper, i.e., $R_{e} \triangleq \frac{1}{n} H(W^k|Y^n_{W})$.

Our goal is to maximize both the transmission rate between Alice and Bob in addition to Eve's uncertainty about the message (equivocation rate). A relaxed secrecy condition is letting $\frac{1}{n} H(W^k)-R_{e} \leq \epsilon$ and $P_{e}^n \leq \epsilon$, where $\epsilon \to 0$ as $n \to \infty$. The secrecy capacity is defined as the maximum achievable secrecy rate for all sequences of $(2^{nR_{s}},n)$ codes 
\begin{equation}
\label{eqn_1}
C_{s} \triangleq \sup\limits_{P_{e}^n \leq \epsilon} R_{s}.
\end{equation}  
The secrecy capacity for AWGN channels (or fixed $\gamma_{M}$ and $\gamma_{W}$ realizations) is given by
\begin{equation}
\label{eqn_2}
C_{s} = \big\{\log(1+\gamma_{M})-\log(1+\gamma_{W})\big\}^{+},
\end{equation}
where $\{x\}^{+} = \max\{x, 0\}$. Thus, the AWGN secrecy capacity is given by the difference between legitimate and eavesdropper channel capacities. By defining the outage probability as $P_{out} = P(C_{s} \leq R_{s})$, the $\epsilon$-outage secrecy capacity is the value of $R_{s}$ that satisfies $P_{out} = \epsilon$.

\section{Secrecy Capacity of Transmission Schemes under Study}
\subsection{Conventional scheme}
In this scheme, all parties are using single antennas and the channel is a block fading channel. Assuming that the legitimate and eavesdropper channel realizations are $\gamma_{M}$ and $\gamma_{W}$ respectively, the secrecy capacity for one realization of both channels is given by (\ref{eqn_2}). For a quasi-static fading channel, we characterize the performance via outage secrecy capacity and outage probability. The outage probability is $P_{out}(R_{s}) = P(C_{s}<R_{s})$, and can be written as
\[
P_{out}(R_{s}) = P(C_{s}<R_{s}|\gamma_{M} > \gamma_{W}) P(\gamma_{M} > \gamma_{W}) 
\]
\[+ P(C_{s}<R_{s}|\gamma_{M} \leq \gamma_{W}) P(\gamma_{M} \leq \gamma_{W}).
\]
From \cite{IEEEhowto:kopka2} we know that $P(\gamma_{M} > \gamma_{W}) = \frac{\overline{\gamma}_{M}}{\overline{\gamma}_{M}+\overline{\gamma}_{W}}$, $P(C_{s}<R_{s}|\gamma_{M} > \gamma_{W}) = 1-\frac{\overline{\gamma}_{M} + \overline{\gamma}_{W}}{\overline{\gamma}_{M} + 2^{R_{s}}\overline{\gamma}_{W}} e^{-\frac{2^{R_{s}}-1}{\overline{\gamma}_{M}}}$, and $P(C_{s}<R_{s}|\gamma_{M} \leq \gamma_{W}) = 1$. Thus, the $\epsilon$-outage secrecy capacity is the value of $R_{s}$ that sets the outage probability to $\epsilon$. The $\epsilon$-outage secrecy capacity $R_{s}$ is obtained by solving the transcendental equation $(1-\epsilon) \big(1+\frac{\overline{\gamma}_{W}}{\overline{\gamma}_{M}} 2^{R_{s}}\big) = e^{-\frac{2^{R_{s}}-1}{\overline{\gamma}_{M}}},$ which can be put in a closed-form in terms of the \textit{Lambert W function}. Thus, the $\epsilon$-outage secrecy capacity is given by
\begin{equation}
\label{eqn_3}
R_{s} = \overline{\gamma}_{M} \times \bigg\{\mathcal{W}_{o}\bigg(\frac{e^{\frac{1}{\overline{\gamma}_{M}}+\frac{1}{\overline{\gamma}_{W}}}}{\overline{\gamma}_{W}(1-\epsilon)}\bigg)  - \frac{1}{\overline{\gamma}_{W}} \bigg\}^{+},
\end{equation}
where $\mathcal{W}_{o}(x)$ is the single valued Lambert W function.

\subsection{State switching scheme (Reconfigurable antennas without CSI)}
In this scheme, the legitimate transmitter and receiver utilize reconfigurable antennas with $Q_{T}$ and $Q_{R}$ radiation states respectively. We assume large codeword lengths and that both $Q_{T}$ and $Q_{R}$ are comprable to $n$. The legitimate channel has $Q_{T}Q_{R}$ possible independent realizations per codeword, each realization correspond to a certain transmitter-receiver antenna state selection. We switch the antenna states such that the channel realization changes every symbol within a codeword. A codeword of length $n$ artificially experiences $n$ coherence intervals as long as $n < Q_{T}Q_{R}$. Thus, the legitimate channel capacity $C_{M}$ for specific $Q_{T}Q_{R}$ legitimate channel realizations with $Q_{T}Q_{R} > n$ is given by $C_{M} = \frac{1}{n} \sum_{i=1}^{n} \log(1+\gamma_{M}^{n}(i))$, for $n \to \infty$ and invoking the \textit{law of large numbers}, we have $C_{M} = E\{\log(1+\gamma_{M})\}$ where $\gamma_{M}$ is an exponential random variable with an average of $\overline{\gamma}_{M}$. By averaging the legitimate channel capacity over the exponential pdf, the legitimate channel can be defined by an \textit{ergodic capacity} as $C_{M} = e^{\frac{1}{\overline{\gamma}_{M}}} \operatorname{Ei}\bigg(\frac{1}{\overline{\gamma}_{M}}\bigg)$ \cite{IEEEhowto:kopka1}, where $\operatorname{Ei}(x) = - \int_{-x}^{\infty}\frac{e^{-t}}{t} dt$ is the exponential integral function. Similarly, the eavesdropper channel has $Q_{T}$ possible channel realizations, and the eavesdropper channel capacity can be written as $C_{W} = e^{\frac{1}{\overline{\gamma}_{W}}} \operatorname{Ei}\bigg(\frac{1}{\overline{\gamma}_{W}}\bigg)$. Therefore, recalling (\ref{eqn_2}), the ergodic secrecy capacity is given by \
\begin{equation}
\label{eqn_5}
C_{s} = \Bigg\{e^{\frac{1}{\overline{\gamma}_{M}}} \operatorname{Ei}\bigg(\frac{1}{\overline{\gamma}_{M}}\bigg) - e^{\frac{1}{\overline{\gamma}_{W}}} \operatorname{Ei}\bigg(\frac{1}{\overline{\gamma}_{W}}\bigg)\Bigg\}^{+}.
\end{equation}
  
\begin{figure}[!h]
\centering
\includegraphics[width=5 in]{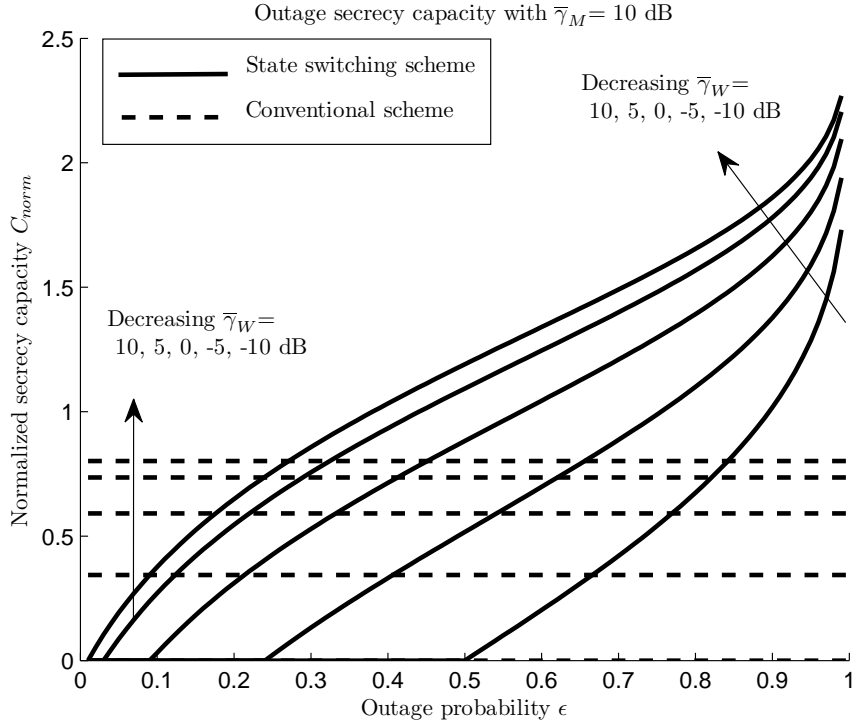}
\caption{Outage capacity of the conventional scheme versus the ergodic capacity of the state switching scheme.}
\label{fig_sim}
\end{figure}

Note that the ergodic definition for the secrecy capacity in (\ref{eqn_5}) describes two artificial fast fading legitimate and eavesdropper channels. Although the wireless channel is quasi-static, reconfigurable antennas can be used to induce channel fluctuations over time by switching the radiation states to emulate fast fading. We are interested in studying whether the ergodic capacity in (\ref{eqn_5}) can be larger than the $\epsilon$-outage secrecy capacity. Figure 2 depicts the outage secrecy capacity (solid lines) plotted versus $\epsilon$ for $\overline{\gamma}_{M}$ = 10 dB together with the ergodic capacity (dashed lines) for different eavesdropper channel average SNR values. Note that the ergodic capacity exceeds the outage capacity for tight outage constraints (i.e. small values of $\epsilon$). For instance, when $\overline{\gamma}_{W}$ = -10 dB, the state switching scheme outperforms the conventional scheme as long as $\epsilon < 0.25$. Moreover, for $\overline{\gamma}_{W}$ = 5 dB, the state switching scheme is better for $\epsilon < 0.4$. Besides, the outage capacity does not exist for $\epsilon < 0.25$, thus the state switching scheme is definitely beneficial for strict outage constraints. On the other hand, when $\overline{\gamma}_{M} \leq \overline{\gamma}_{W}$, the ergodic capacity does not exist and only outage capacity with relaxed outage probability constraints is realizable.              

\subsection{State selection with partial CSI}

In this scheme, the legitimate transmitter has the legitimate channel CSI but is not provided with the eavesdropper channel CSI. Thus, the transmitter and the receiver can agree on the adopted radiation states at both parties once per codeword. The secrecy capacity is given by 
\[
C_{s} = \sup\limits_{1 \leq j \leq Q_{T}Q_{R}} \Bigg\{\log(1+\gamma_{M, j}) - \log(1+\gamma_{W})\Bigg\}^{+},
\]
where $\gamma_{M, j}$, with $j \in \{1, 2, \ldots, Q_{R}Q_{T}\}$, is one of the $Q_{R}Q_{T}$ independent identical channel realizations obtained by different combinations of the transmission and reception radiation states \cite{IEEEhowto:kopka5}, and $\gamma_{W}$ is a Rayleigh random variable and represents the corresponding eavesdropper channel obtained from a certain selection of the radiation states. Intuitively, the secrecy capacity for a certain set of $Q_{R}Q_{T}$ channel realizations is given by
\begin{equation}
\label{eqn_7}   
C_{s} = \Bigg\{\log(1+\gamma_{M, max}) - \log(1+\gamma_{W})\Bigg\}^{+},
\end{equation}
where $\gamma_{M, max} = \max\{\gamma_{M, 1}, \gamma_{M, 2}, \ldots, \gamma_{M, Q_{R}Q_{T}}\}$. The pdf of $\gamma_{M, max}$ is \cite{IEEEhowto:kopka7}
\[f_{\gamma}(\gamma_{M, max}) = Q_{T}Q_{R} \sum_{i=0}^{Q_{T}Q_{R}} \binom{Q_{T}Q_{R}-1}{i} \frac{(-1)^i}{\overline{\gamma}_{M}} e^{-\frac{\gamma_{M, max}}{\overline{\gamma}_{M}/(i+1)}},\] 
thus, it can be easily shown that $P(\gamma_{M, max} > \gamma_{W}) = Q_{T}Q_{R} \sum_{i=0}^{Q_{T}Q_{R}} \binom{Q_{T}Q_{R}-1}{i} \frac{(-1)^i}{i+1} \frac{1}{1+\frac{(i+1) \overline{\gamma}_{W}}{\overline{\gamma}_{M}}}$, whereas $P(C_{s}<R_{s} | \gamma_{M, max}>\gamma_{W})$ is given in (\ref{eqn_dbl_x1}). As demonstrated before, $P_{out}(R_{s}) = P(C_{s}<R_{s}|\gamma_{M, max} > \gamma_{W}) P(\gamma_{M, max} > \gamma_{W})+ P(C_{s}<R_{s}|\gamma_{M, max} \leq \gamma_{W}) P(\gamma_{M, max} \leq \gamma_{W})$ and the $\epsilon$-outage secrecy capacity is obtained by solving the transcendental equation in (\ref{eqn_dbl_x2}) for $R_{s}$.

\begin{equation}
\label{eqn_dbl_x1}
P(C_{s}<R_{s} | \gamma_{M, max}>\gamma_{W}) = Q_{T}Q_{R} \sum_{i=0}^{Q_{T}Q_{R}} \binom{Q_{T}Q_{R}-1}{i} \frac{(-1)^i}{(i+1)} \Bigg( \frac{1}{1+\frac{(i+1)\overline{\gamma}_{W}}{\overline{\gamma}_{M}}} - \frac{e^{\frac{-(2^{R_{s}}-1)(i+1)}{\overline{\gamma}_{M}}}}{1+\frac{(i+1) 2^{R_{s}}\overline{\gamma}_{W}}{\overline{\gamma}_{M}}}\Bigg).
\end{equation}

\[\frac{1-\epsilon}{\Bigg(Q_{T}Q_{R} \sum_{i=0}^{Q_{T}Q_{R}} \binom{Q_{T}Q_{R}-1}{i} \frac{(-1)^i}{i+1} \frac{1}{1+\frac{(i+1) \overline{\gamma}_{W}}{\overline{\gamma}_{M}}}\Bigg)}=\]
\begin{equation}
\label{eqn_dbl_x2}
\Bigg(1- Q_{T}Q_{R} \sum_{i=0}^{Q_{T}Q_{R}} \binom{Q_{T}Q_{R}-1}{i} \frac{(-1)^i}{(i+1)} \Bigg( \frac{1}{1+\frac{(i+1)\overline{\gamma}_{W}}{\overline{\gamma}_{M}}} - \frac{e^{\frac{-(2^{R_{s}}-1)(i+1)}{\overline{\gamma}_{M}}}}{1+\frac{(i+1) 2^{R_{s}}\overline{\gamma}_{W}}{\overline{\gamma}_{M}}}\Bigg)\Bigg).
\end{equation}

\subsection{State selection with full CSI}
Assume that both the legitimate and the eavesdropper channel CSI are available at the legitimate parties. In this case, state selection will be applied such that the legitimate channel is maximized while the eavesdropper channel is minimized. Because the legitimate channel depends on the selection of one of $Q_{T}$ transmitter radiation states, and one of $Q_{R}$ receiver radiation states, we have a total of $Q_{T}Q_{R}$ possible independent channel realizations. On the other hand, the eavesdropper channel depends only on the transmitter radiation state and thus has one of $Q_{T}$ possible channel realizations. Let the legitimate channel be denoted by $\gamma_{M}^{i,j}$ where $i \in \{1,\ldots,Q_{T}\}$ and denotes the selected transmitter radiation state, whereas $j \in \{1,\ldots,Q_{R}\}$ and denotes the selected receiver state. Similarly, the eavesdropper channel is $\gamma_{W}^{i}$ where $i \in \{1,\ldots,Q_{T}\}$, thus we note that the selection of a transmitter radiation state dictates an eavesdropper channel and a set of possible $Q_{R}$ legitimate channels, from where a single realization is picked based on the receiver state. The achievable secrecy capacity for a certain set of legitimate and eavesdropper channel realizations corresponds to the supremum of all selections for transmitter and receiver radiation states
\begin{equation}
\label{eqn_8}
C_{s} = \sup\limits_{1 \leq i \leq Q_{T}, 1 \leq j \leq Q_{R}} \Bigg\{\log(1+\gamma_{M}^{i,j}) - \log(1+\gamma_{W}^{i})\Bigg\}^{+}.
\end{equation}       
Equation (\ref{eqn_8}) suggests that we do not only improve the legitimate channel, but also use the CSI to select the radiation state that undermines the eavesdropper channel. Numerical results for the $\epsilon$-outage secrecy capacity $R_{s}$ are obtained in section IV.

\section{Numerical results}
In figure 3, we investigate the achievable $\epsilon$-outage secrecy capacity for different schemes. By setting $\epsilon$ = 0.1, we plot the secrecy capacity versus $\overline{\gamma}_{M}$ for low and high values of $\overline{\gamma}_{W}$ (-10 and 20 dB respectively). Note that adopting single antennas (conventional scheme) causes an SNR loss of around 10 dB for $\overline{\gamma}_{W}$ = 20 and -10 dB compared to the AWGN secrecy capacity. Reconfigurable antennas provide considerable SNR gains when the CSI is available. Without CSI, the state switching scheme provides poor ergodic capacity for $\overline{\gamma}_{W}$ = 20 dB. For $\overline{\gamma}_{W}$ = -10 dB, the ergodic capacity of the state switching scheme outperforms the outage capacity of the conventional single antenna system for $\overline{\gamma}_{M} <$ 25 dB. In this case, an SNR gain of about 5 dB is achieved. Thus, the state switching scheme offers an SNR gain only for low values of $\overline{\gamma}_{M}$. On the other hand, the state selection with partial CSI scheme offers a significant gain for all legitimate and eavesdropper SNR ranges. The number of radiation states involved in calculations are $Q_{T}$ = $Q_{R}$ =5. For $\overline{\gamma}_{W}$ = -10 dB, partial CSI offer 5 dB SNR gain compared to the AWGN capacity and 15 dB compared to the single antenna system in Rayleigh fading. For high eavesdropper average SNR ($\overline{\gamma}_{W}$ = 20 dB), an SNR gain of 2 dB compared to the AWGN capacity and 12 dB compared to the conventional scheme. It is worth mentioning that the achievable gain is higher for lower values of $\overline{\gamma}_{W}$ as this scheme is not provided with the eavesdropper channel CSI. Moreover, the state selection scheme with full CSI provide superior secrecy capacity compared to all other schemes. This gain is most notable for large $\overline{\gamma}_{W}$, i.e. for $\overline{\gamma}_{W}$ = 20 dB, where and SNR gain of 20 dB compared to the AWGN capacity and 30 dB compared to the single antenna fading channel capacity. The reason for such impressive performance boost is that knowledge of the eavesdropper CSI and the selection of the ``worst" eavesdropper channel is most effective when the eavesdropper channel enjoys high SNR. The gain achieved for $\overline{\gamma}_{W}$ = -10 dB is about 8 dB compared to the AWGN channel. The gain decreases for low values of $\overline{\gamma}_{W}$ because undermining the eavesdropper channel becomes of less effectivness. 

Figure 4 demonstrates the secrecy capacity normalized to the AWGN secrecy capacity for $\epsilon$ = 0.1 and $\overline{\gamma}_{W}$ = -10, 0 and 10 dB. Focusing on the state selection scheme with full CSI, we note that the capacity gain is maximum at low SNR. This is similar to the effect of optimal water filling power allocation over a fast fading channel, where the highest capacity gain is obtained at low SNR. In our case power allocation is applied across the radiation state domain rather than the time domain. Besides, instead of water filling, we allocate all power to the best radiation state. Thus, we are able to achieve considerable capacity gains regardless of the time latency of the applications, i.e. delay tolerant and delay sensitive applications are both supported by the reconfigurable antenna scheme.              
 
\begin{figure}[!h]
\centering
\includegraphics[width=5in]{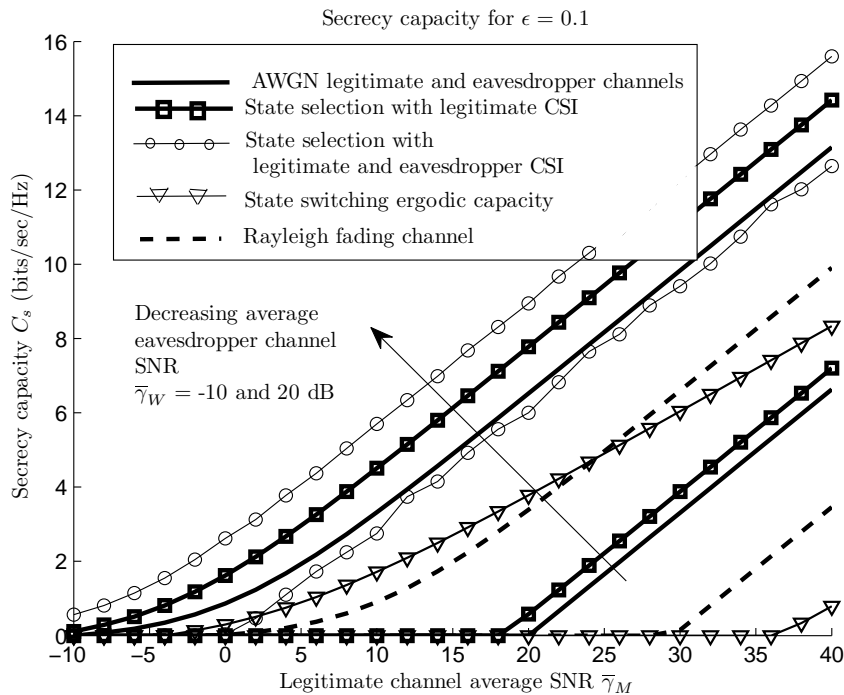}
\caption{Comparison between all secret communications schemes.}
\label{fig_sim}
\end{figure}

\begin{figure}[!h]
\centering
\includegraphics[width=5in]{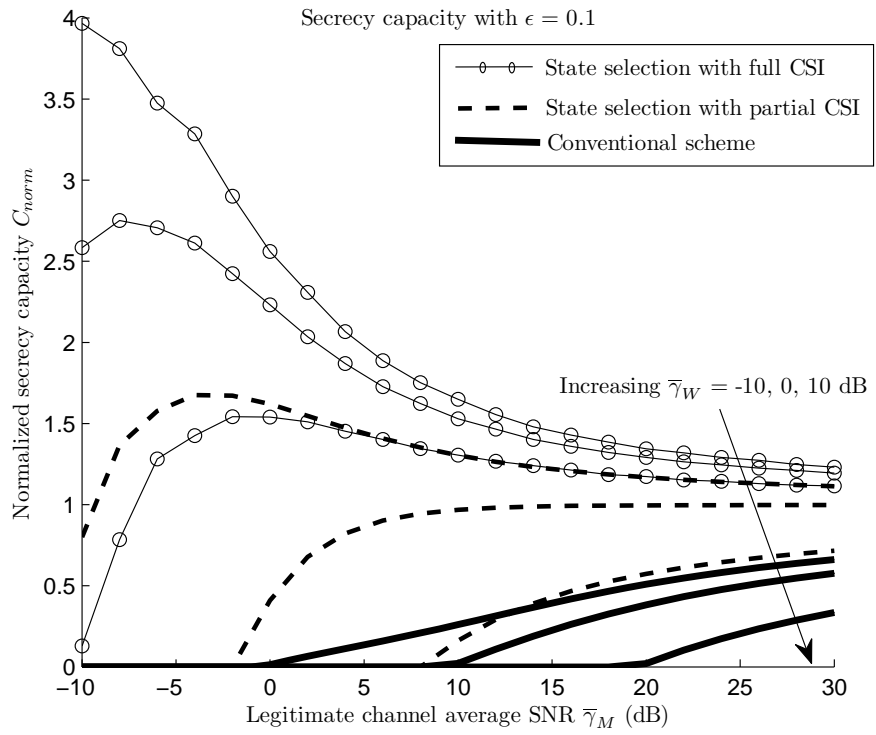}
\caption{Secrecy capacity gain for different schemes.}
\label{fig_sim}
\end{figure}


\end{document}